\documentstyle[11pt,paspconf,epsf]{article}
\begin{document}
%======================
{\scriptsize \noindent
Invited review to be published in ``Abundance Profiles: Diagnostic Tools 
for Galaxy History''
\\ eds. D. Friedli, M.G. Edmunds, C. Robert, \& L. Drissen, ASP Conf. Ser.
\vskip -1truemm}
%======================
%----------------------------------------------------------------------
\title{Mixing and Transfer of Elements by Triaxiality} 
\author{Daniel Friedli} 
\affil{D\'epartement de Physique and Observatoire du Mont M\'egantic,\\
Universit\'e Laval, Qu\'ebec, Qc, G1K~7P4, Canada; dfriedli@phy.ulaval.ca
}

%----------------------------------------------------------------------
\begin{abstract}
Various large-scale processes induced by triaxial potentials and able
to alter stellar and gaseous abundance profiles are reviewed.  In
particular, strong bars can quickly and efficiently blend stars
through chaotic mixing, and entail intricate inflows/outflows of gas.
Bar-generat\-ed abundance features include flattenings, breaks, and
plateaus in radial gradients as well as box- or peanut-like
morphologies.
\end{abstract}

%----------------------------------------------------------------------
% KEYWORDS SHOULD BE INCLUDED, BUT THEY ARE NOT PRINTED IN THE HARDCOPY!

%----------------------------------------------------------------------
\section{Introduction}
In the world of galaxies, triaxial structures are very common.
Fast-rotating ones include stellar bars which are present in at least
two thirds of disk galaxies (Sellwood \& Wilkinson 1993), whereas
nearly 100\% of Magellanic-type galaxies host bars (Odewahn 1996).
Gaseous bars might also be present in primordial disks.  Slow- or
non-rotating triaxial potentials probably include bright ellipticals,
but not faint ones (Merritt \& Valluri 1996).  Triaxial systems have
at least two outstanding properties: they may contain irregular
(chaotic) orbits, and they induce gravitational torques. These
characteristics lead both to large-scale diffusion and mixing of
stars, and produce transfer of angular momentum and matter; here,
large-scale means $\ga 1$\,kpc. For mechanisms involved at smaller
scales, see for instance Roy \& Kunth (1995) and Elmegreen (this
volume).

Stellar and gaseous components have distinct dynamics.  The stellar
one is mainly affected by large-scale mixing processes like {\sl
chaotic mixing}, which is generated by irregular orbits in triaxial
potentials. Other mixing processes include the well-known {\sl phase
mixing} which operates e.g. through differential rotation, and {\sl
violent relaxation} which is relevant in strongly time-dependent
potentials as present during galaxy formation and mergings (Binney \&
Tremaine 1987).  Diffusion by encounters is also important but will
not be discussed here (see e.g. Binney \& Lacey 1988).  On the
contrary, the cold or warm gaseous components are essentially affected
by transfer processes, which operate via {\sl gravitational}, {\sl
viscous}, or {\sl magnetic torques}.  At kpc scales in strongly
non-axisymmetric potentials, gravitational torques generally dominate
over the two others.

Stellar abundances are becoming available for external galaxies,
either from individual stars (Monteverde et al. 1997), or from
integrated light (Beauchamp 1997). Numerical models presented in
Sect.~2.2 thus provide predictions which still have to be confirmed by
observations.  On the contrary, numerous abundance determinations
exist for the gas phase (Henry, this volume), and numerical results
can be compared to observations (Sect.~3.2).

%----------------------------------------------------------------------
\section{Mixing and Transfer of Stars}
%----------------------------------
\subsection{Chaos in Triaxial Systems}
%----------------------------------
\subsubsection{Barred Galaxies.}
Many investigations have been dedicated to the study of stellar orbits
in barred potentials (e.g. reviews by Contopoulos \& Grosb\/{\o}l
1989; Sellwood \& Wilkinson 1993). For these systems, one of the most
remarkable discoveries is the possible existence of irregular orbits
able to tour the whole phase space in less than a Hubble time
$\tau_{\rm H}$.  These chaotic orbits are present around the main
resonances, especially the corotation resonance (CR).  Their fraction
increases when bar strength (Athanassoula et al. 1983), central mass
concentration (Hasan \& Norman 1990), noise or asymmetries (Habid et
al. 1997) increase.  In a recent study on the vertical orbital
structure around CR, Oll\'e \& Pfenniger (1998) have found that
significant and fast radial and vertical diffusions take place when a
critical bar strength is reached.  The reason is that the Lagrangian
points $L_{4,5}$ are then becoming ``complex unstable''.  The analysis
of the population of self-consistent quasi-stable N-body strong bars,
in 2D (Sparke \& Sellwood 1987) or 3D (Pfenniger \& Friedli 1991), has
revealed that typically $\sim 35\%$ of orbits are chaotic ({\sl hot
population}), $\sim 45\%$ are confined inside the bar region ({\sl
bar population}), and $\sim 20\%$ are in the disk region ({\sl disk
population}).  The high number of chaotic orbits clearly suggests
that chaotic mixing should be important in such galaxies and might
have dramatic consequences on stellar abundance profiles as will be
discussed in Sect.~2.2.

%----------------------------------
\subsubsection{Elliptical Galaxies.}
Several orbit studies connected with the onset of chaos in triaxial
potentials, similar to real elliptical galaxies, have been undertaken
(e.g. Udry \& Pfenniger 1988; Martinet \& Udry 1990; Schwarzschild
1993; Merritt \& Fridman 1996).  As for barred galaxies, chaos
increases when triaxiality, central mass concentration, noise or
asymmetries increase.  Merritt \& Valluri (1996) performed a detailed
study on chaos and mixing in realistic potentials for
ellipticals. They found that potentials with central density cusps
present chaotic mixing timescales smaller than $\tau_{\rm H}$,
especially near the center.  As a result, triaxial galaxy centers
should become axisymmetric with time.  To my knowledge, the
consequences on abundance profiles have not been studied so far; one
could however expect a central flattening of abundance gradients,
which is apparently not observed (Worthey, this volume).

%----------------------------------
\subsection{Bars: Consequences for Stellar Abundance Profiles}
N-body simulations represent a powerful tool to investigate non-linear
phenomena, like those involving bar formation and evolution.
Following previous studies of bar effects on abundance profiles
(Friedli et al. 1994; Friedli \& Benz 1995), I recently performed a
new series of collisionless simulations of spontaneous bar formation
(no element production); a particle-mesh method with $N=400\,000$, a 3D
exponential polar grid ($N_R\!=\!61$, $N_{\phi}\!=\!64$,
$N_z\!=\!243$), and a time-step $\Delta t \!=\! 0.1 \, \rm Myr$ are
used.  The central radial resolution is $20 \, \rm pc$, the vertical
one $100 \, \rm pc$.  Initial models are axisymmetric with
pre-existing exponential abundance gradients, i.e.:
\begin{equation}
{d\log ({\rm A/H}) \over dX} \equiv {\rm A}_{X}^{\rm reg} = {\rm constant}
\hskip 5truemm
[\rm dex \, kpc^{-1}] \, , 
\end{equation}
where A stands for the chemical species studied (e.g. O for oxygen),
$X$ for radial $R$, azimuthal $\phi$, or vertical $z$ axes, and
``reg'' for the region considered.

%----------------------------------
\begin{figure}
\plottwo{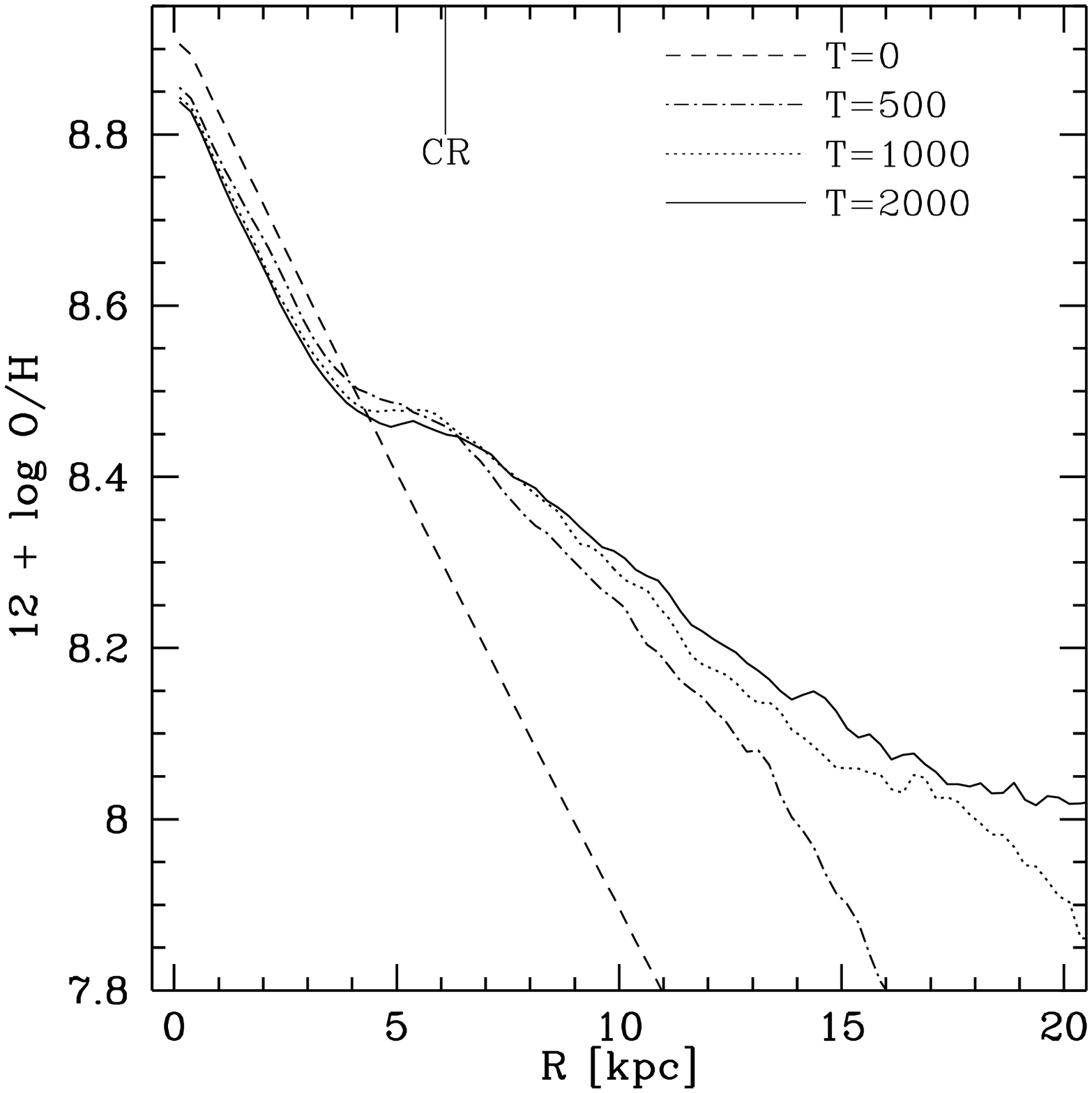}{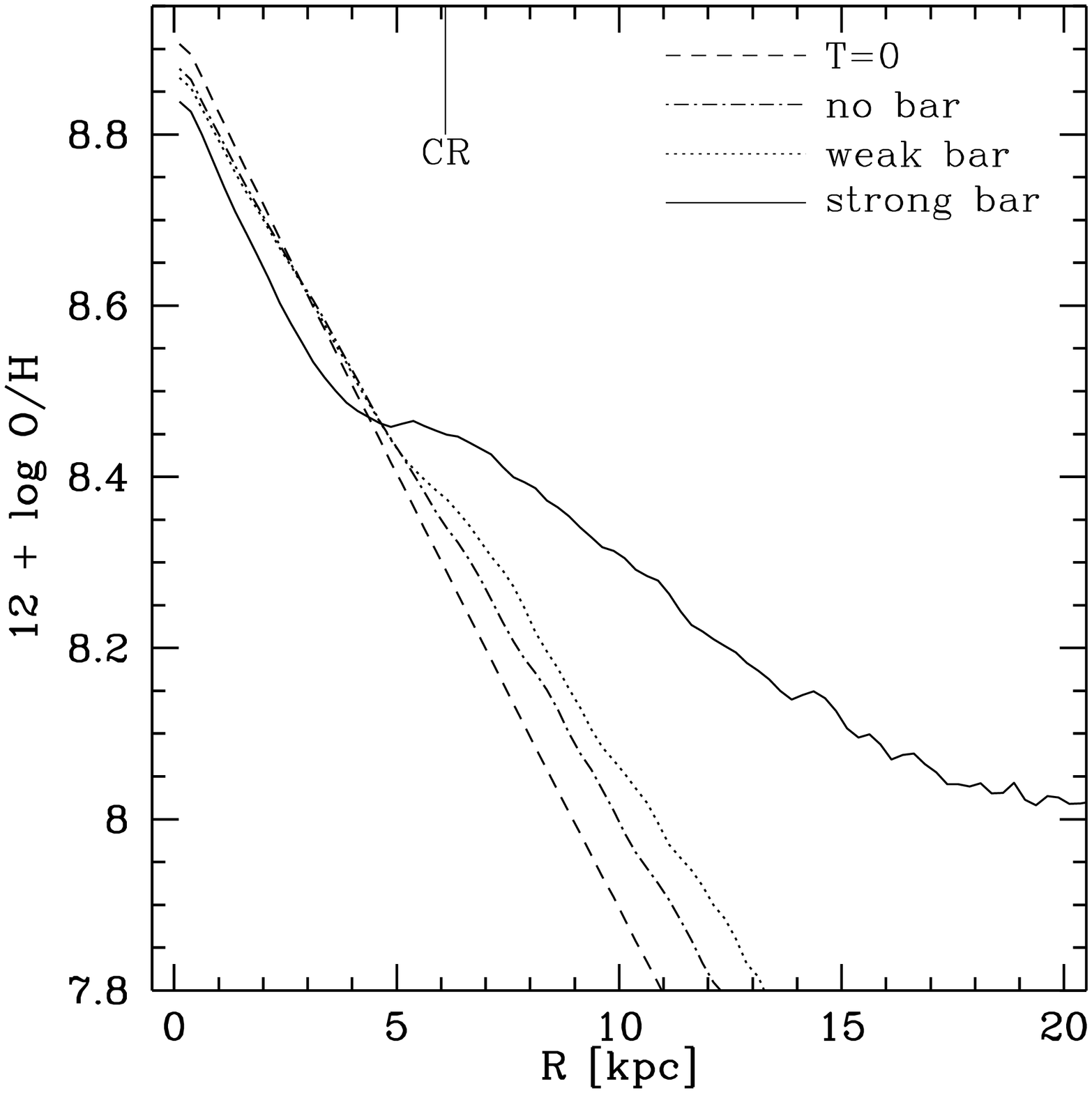}
\caption{{\it Left.}  Time evolution of the stellar radial abundance
profile in a galaxy model forming a strong bar. CR indicates the
corotation radius at $T \!=\! 2000$\,Myr.  {\it Right.}  Comparison of
radial gradients between systems having no bar, a weak bar, or a
strong one ($T \!=\!  2000$\,Myr).}
\end{figure}

%----------------------------------
\subsubsection{Azimuthal Profile.}
The first case studied is ${\rm O}_{\phi}^{\rm gal} \!=\!  -0.5$, and
${\rm O}_{R}^{\rm gal} \!=\!  {\rm O}_{z}^{\rm gal} \!=\! 0.0$.  Here,
phase mixing is the dominant process.  A very quick homogenization of
the azimuthal gradient is observed both in axisymmetric and barred
models, mixing being however $\sim 1.5$ faster in barred models.
Like dynamical timescales, mixing timescales increase with $R$.  After
$\sim 1$\,Gyr, no variations of mean abundances are detected in the
whole disk.

%----------------------------------
\subsubsection{Radial Profile.}
The ``standard'' case has ${\rm O}_{R}^{\rm gal} \!=\!  -0.1$, and
${\rm O}_{\phi}^{\rm gal} \!=\! {\rm O}_{z}^{\rm gal} \!=\!  0.0$.
Figure~1 (left) shows the time evolution of the radial abundance
gradient in a model forming a strong bar (maximum bar axis ratio
$(b/a)_{\rm max} \approx 0.36$).  The shape of this gradient evolves
very quickly.  Roughly speaking, it presents a break near CR; in more
details, three major features can be highlighted: i) in the {\sl disk
region}, the gradient becomes much {\it flatter}, typically by a
factor of 2--3, and the mean metallicity is strongly increased (only
due to the redistribution of elements);  ii) in the {\sl bar region},
the gradient remains approximately the same, but the mean metallicity
is decreased; iii) in the {\sl corotation region}, a distinct {\it
plateau} appears.  Note that high spatial and temporal resolutions are
required to be able to observe such a plateau.  Recent observations of
three barred galaxies by Beauchamp (1997) seem to indicate that the
Mg$_2$ index presents such characteristics. However, it remains to be
demonstrated that these features will be preserved after the
conversion of the index into actual Mg abundances.  Note that the
projected isoabundances appear barred as well (Fig.~3, right).

%To check that the bar is really the main engine which shapes the
%radial abundance gradient, additional systems with no bar ($(b/a)_{\rm
%max} \!=\! 1.00$), and a weak bar ($(b/a)_{\rm max} \approx 0.70$),
%have also been computed. 
To check that the bar is really the main engine which shapes the
radial abundance gradient, additional systems with a weak bar (bar
strength three times smaller than in the strong bar case) and no bar
have also been computed.  Figure~1 (right) compares the gradient of
these various models at $T \!=\! 2000$\,Myr.  Observed changes clearly
appear to be dependent on the bar strength, as is also expected for
chaotic mixing (see Sect. 2.1). The preponderant role of chaos is seen
in Fig.~2 (left), where the respective gradients for the total, bar,
hot, and disk populations have been plotted. The hot particles present
a nearly flat gradient across the whole galaxy. Since they dominate
the density distribution near CR, the origin of the plateau is
disclosed!  Although the mean gradient is flat, a huge abundance
scatter, of the order of 1\,dex, is present (Fig.~2, right).

The Sun is over-abundant with respect to the surrounding stars
(e.g. Meyer et al. 1998). One explanation is that it was formed in a
richer environment a few kpc closer to the galactic center, and then
moved outwards. Since the Milky Way has a bar, a bar-induced diffusion
might have caused this migration.

%----------------------------------
\begin{figure}
\plottwo{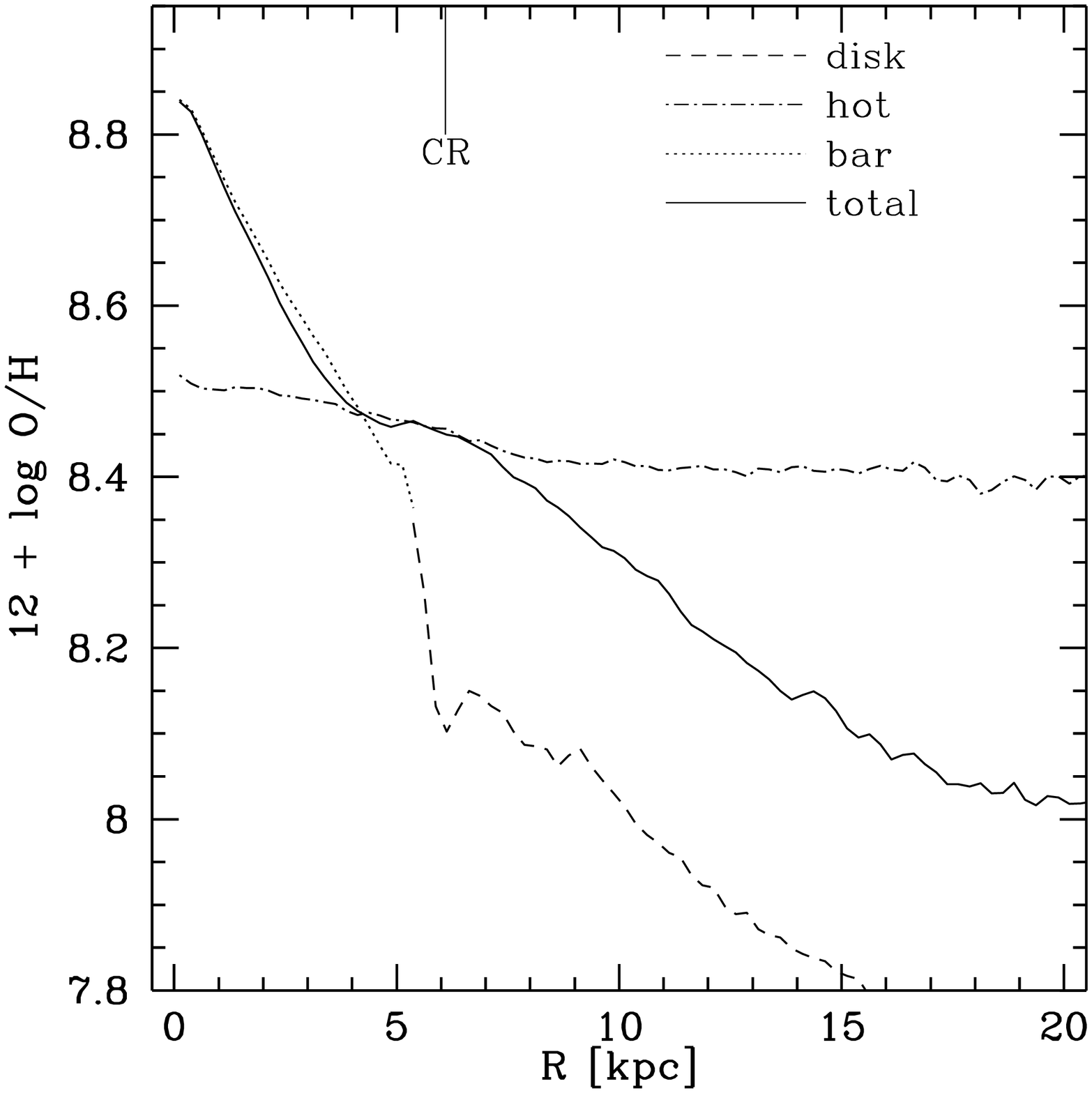}{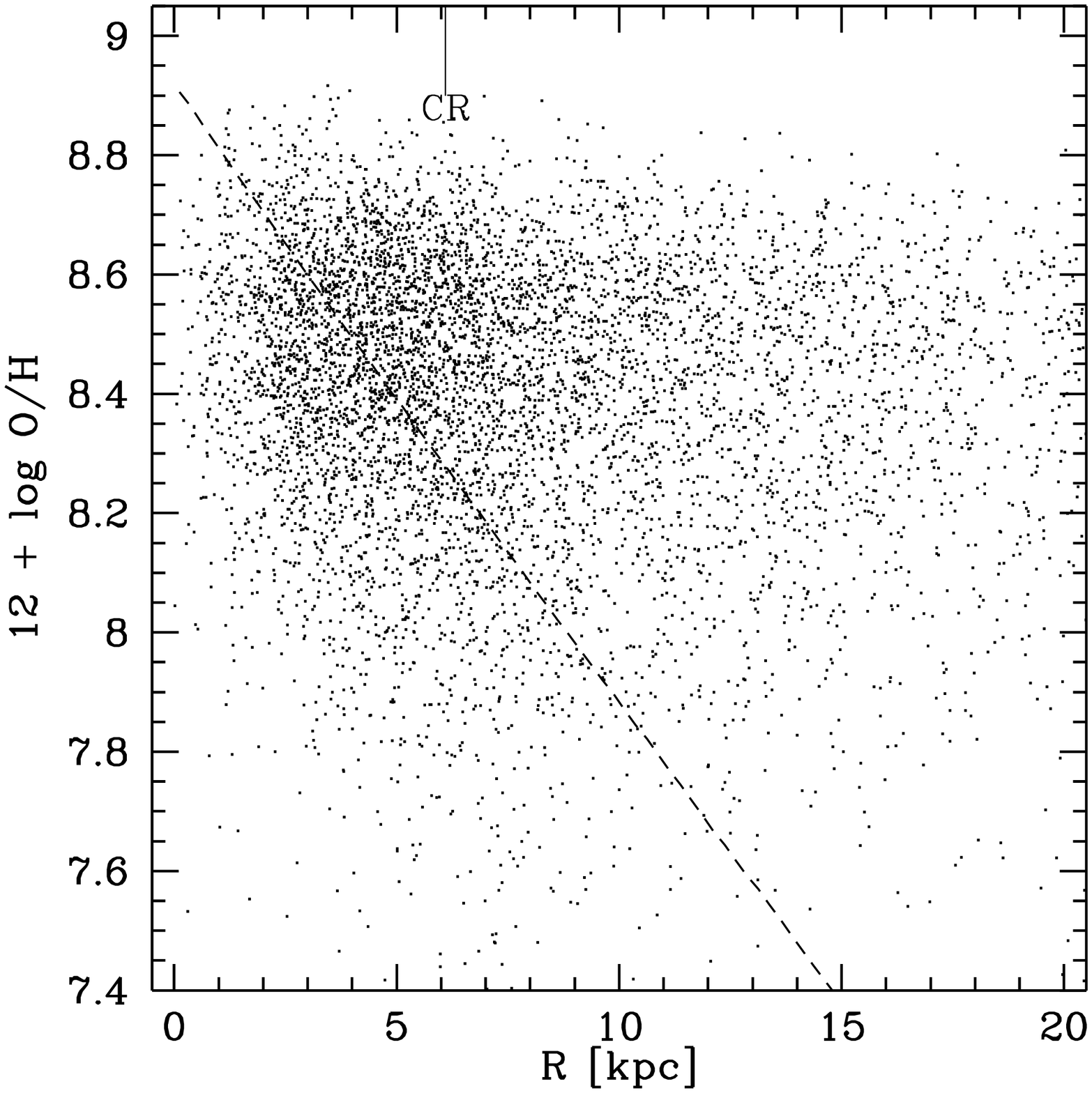}
\caption{{\it Left.}  Stellar radial abundance profiles in the
strongly barred model for the total, bar, hot, and disk population of
particles ($T \!=\! 2000$\,Myr).  CR indicates the corotation radius.
{\it Right.}  Abundance scatter in the hot population. The dashed line
is the initial value. Only 1/20 of the particles are plotted.}
\end{figure}

%----------------------------------
\begin{figure}
\plottwo{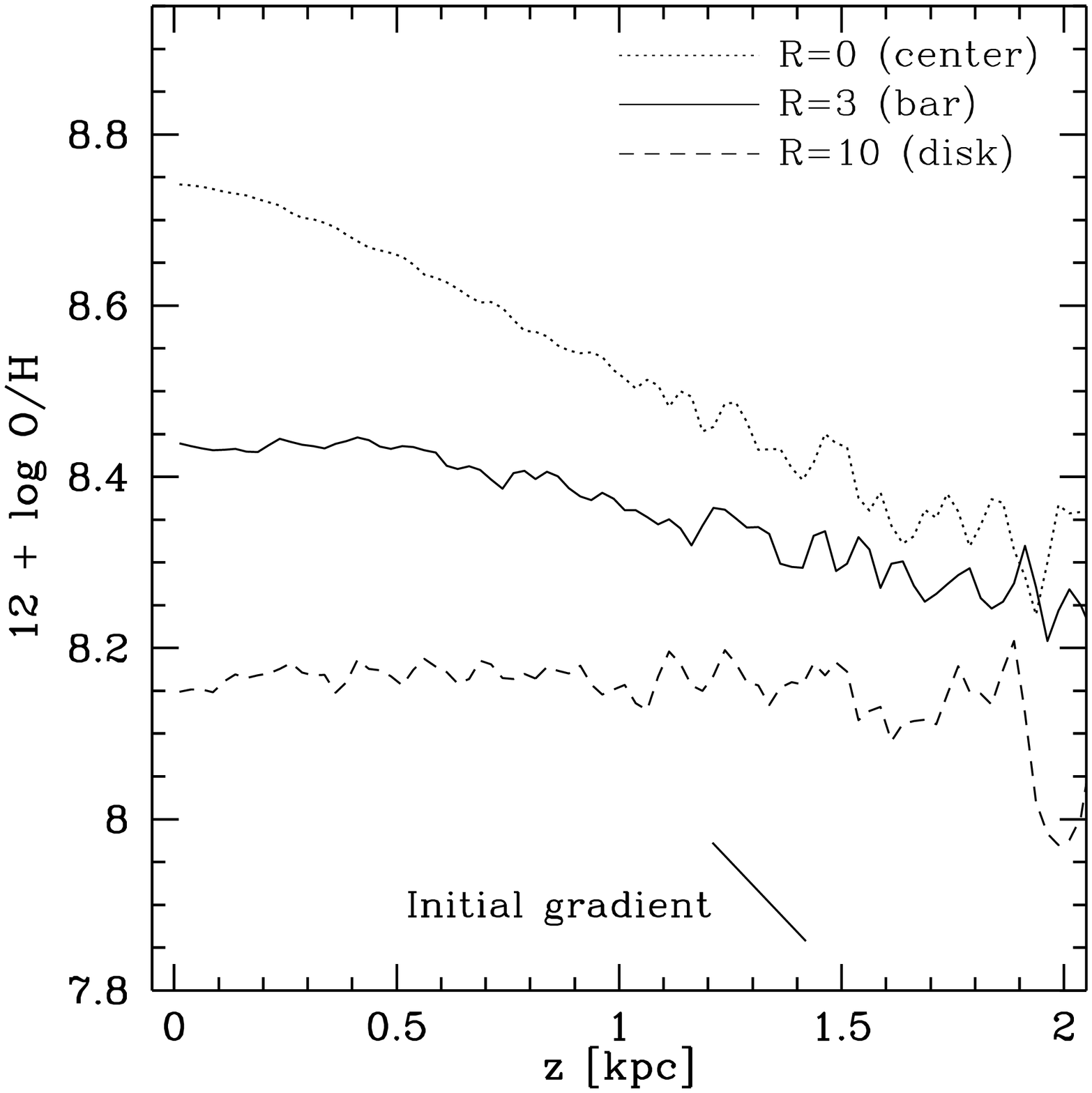}{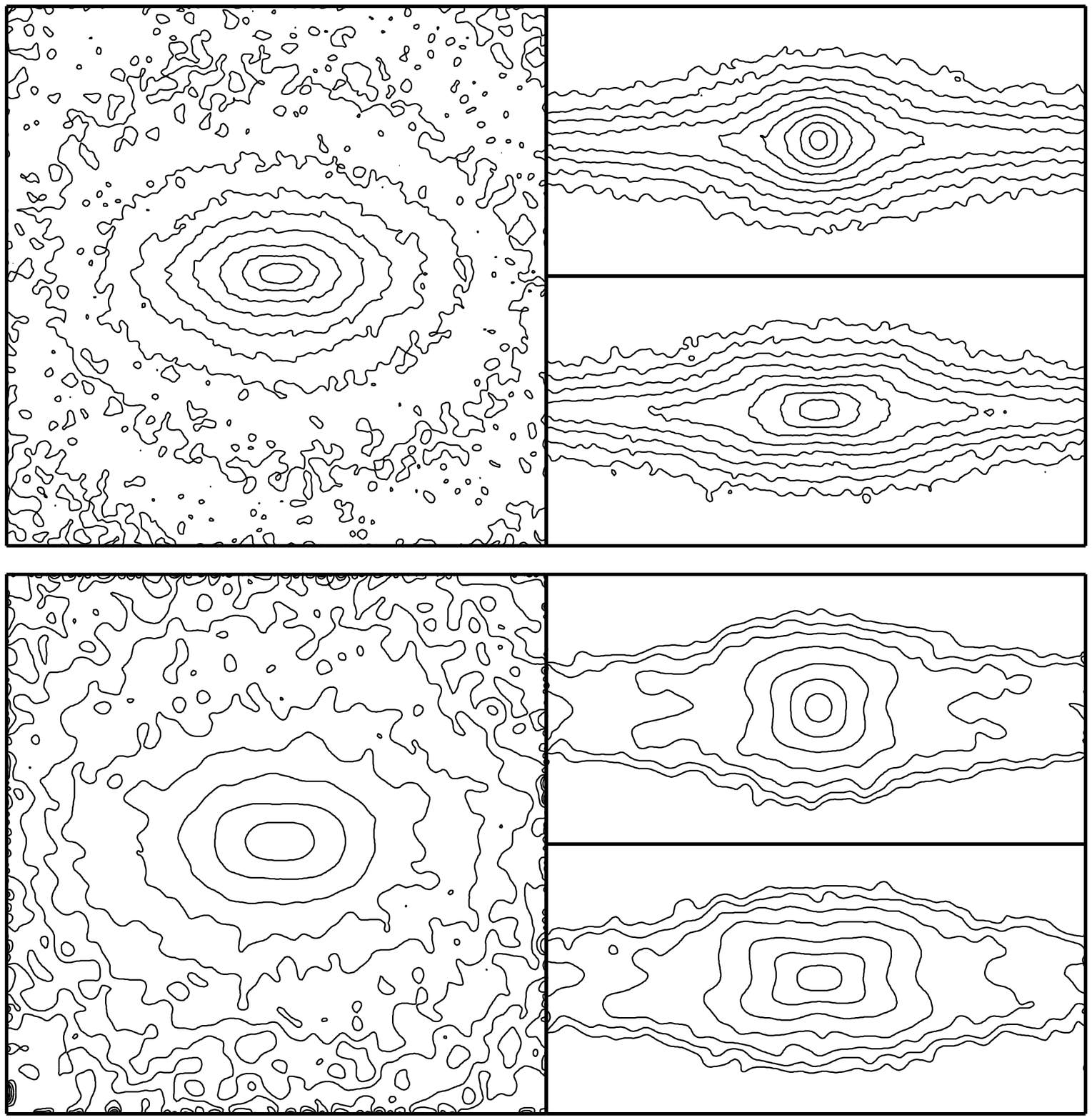}
\caption{{\it Left.}  Stellar vertical abundance gradients at various
radii $R$ in a galaxy model forming a strong bar ($T \!=\!
1000$\,Myr).  {\it Right.}  Projected isodensity (top) and isoabundance
(bottom) stellar contours in the three principal planes ($T \!=\!
2000$\,Myr). The scales are logarithmic and the frames are 8\,kpc
wide. Note the striking boxy morphology.}
\end{figure}

%----------------------------------
\subsubsection{Vertical Profile.}
Finally, let us concentrate on the evolution of vertical abundance
profiles by studying the case ${\rm O}_{z}^{\rm gal} \!=\!  -0.5$,
${\rm O}_{R}^{\rm gal} \!=\! -0.1$, and ${\rm O}_{\phi}^{\rm gal}
\!=\!  0.0$.  Figure~3 (left) shows that the pre-existing steep
vertical gradient is quickly flattened both in the bar and disk
regions. A significant gradient only remains at $R\!=\!0$.  In fact,
projected isoabundances have a distinct boxy shape inside the bar
(Fig.~3, right), more precisely inside the vertical Inner Lindblad
Resonance (ILR).  Bars and their associated vertical resonances have
been recognized as very efficient engines to produce box/peanut shaped
isodensity contours (Pfenniger 1985; Combes et al. 1990; Pfenniger \&
Friedli 1991). And this morphology is even more pronounced when the
abundances are plotted.  In models having ${\rm O}_{z}^{\rm gal}
\!=\!0$ at the beginning, a moderate negative vertical gradient
appears at $R\!=\!0$, and a {\it positive} vertical gradient is
observed in the disk. It results from the $z$- and $R$-diffusion of
the hot metal-rich particles.  The box/peanut shape is also becoming
more extreme to be definitely X-shaped in this case.

%----------------------------------------------------------------------
\section{Mixing and Transfer of Gas}
%----------------------------------
\subsection{Bar-Induced Gas Flows and Star Formation}
If gas moves on intersecting trajectories (e.g. chaotic orbits or
periodic orbits with loops), it will be shocked. Kinetic energy is
first transformed into thermal energy, which in turn is radiated away
out of the galaxy by escaping photons, since radiative cooling is
generally very efficient. In barred galaxies, significant energy
dissipation via shocks, as well as angular momentum transport via
gravitational torques, result in intricate transfers (flows) and
mixing of cold or warm gas. This has been demonstrated by
hydrodynamical simulations (Prendergast 1983; Athanassoula 1992;
Friedli \& Benz 1993; Friedli et al. 1994; Piner et al. 1995).  Very
roughly, one observes: i) net gas inflows inside CR towards the
center, channeled along the bar major axis; ii) outflows outside CR
towards the Outer Lindblad Resonance via bi-symmetric spiral arms.
Inflow/outflow rates are proportional to the bar strength, but they
also depend on other parameters like the gas mass fraction.  While the
effects of ``simple'' radial gas flows can be incorporated in galactic
chemical evolution models (e.g. Edmunds \& Greenhow 1995), such an
approach appears untractable for realistic flows in strongly triaxial
systems, where self-consistent numerical simulations are necessary.

Strictly speaking, star formation (SF), and consequently element
production, are beyond the scope of this review! However, when dealing
with gaseous abundances, it is natural to introduce SF in the models,
which entails important results (Sect.~3.2).  The possible connections
between bars and SF have been widely studied (Hawarden et al. 1986;
Kennicutt 1994; Contini 1996; Martinet \& Friedli 1997; Martin \&
Friedli 1997; and references therein).  One of the main results is
that young ($\la 1$\,Gyr) and strong bars significantly enhance SF.
Also, the sites of massive star formation within strong bars evolve,
over a few Gyr, from an extended chain-like distribution along the bar
major axis and the center, to a concentrated nuclear ring-like
morphology once ILRs appear.

%----------------------------------
\begin{figure}
\plottwo{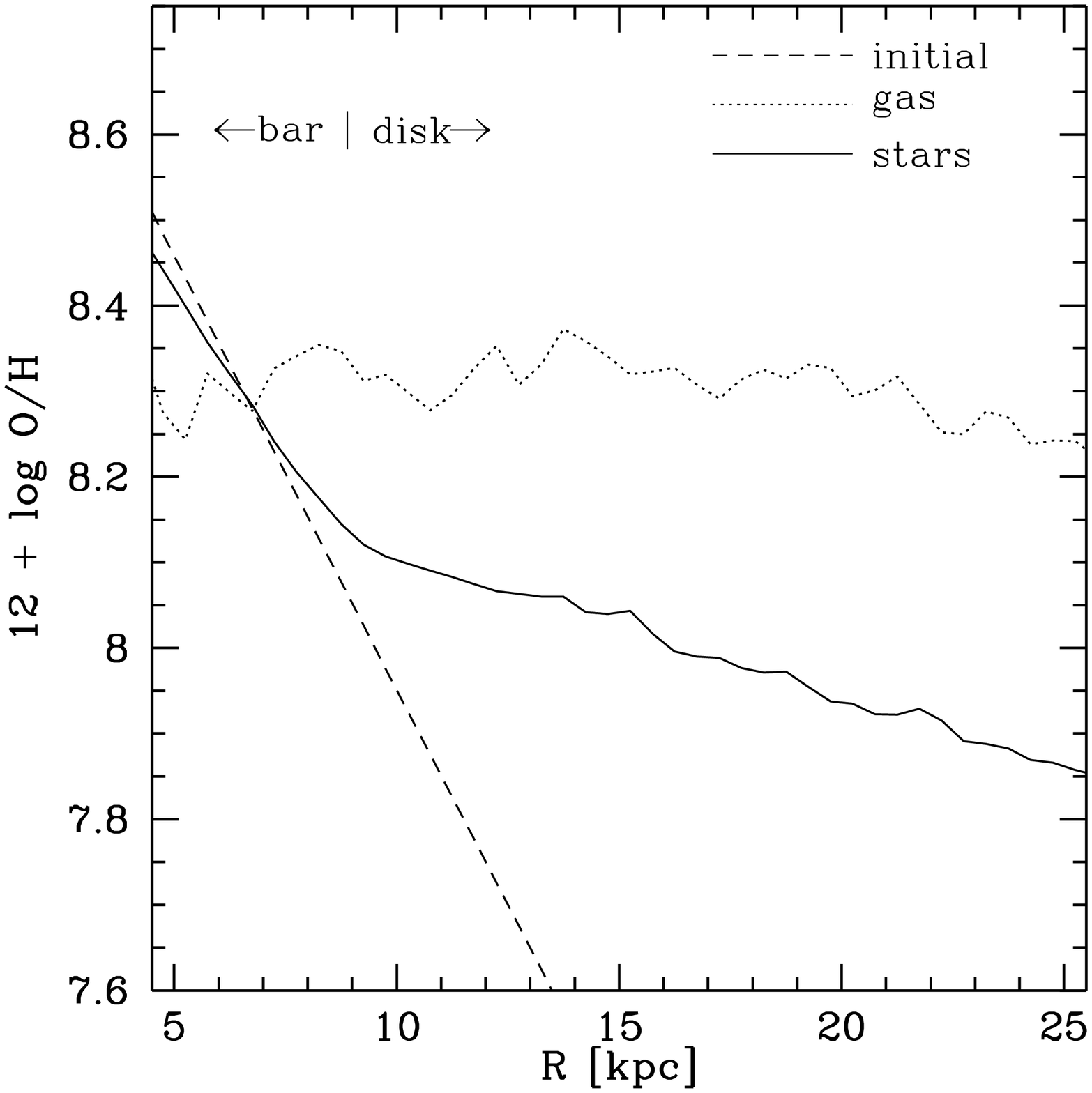}{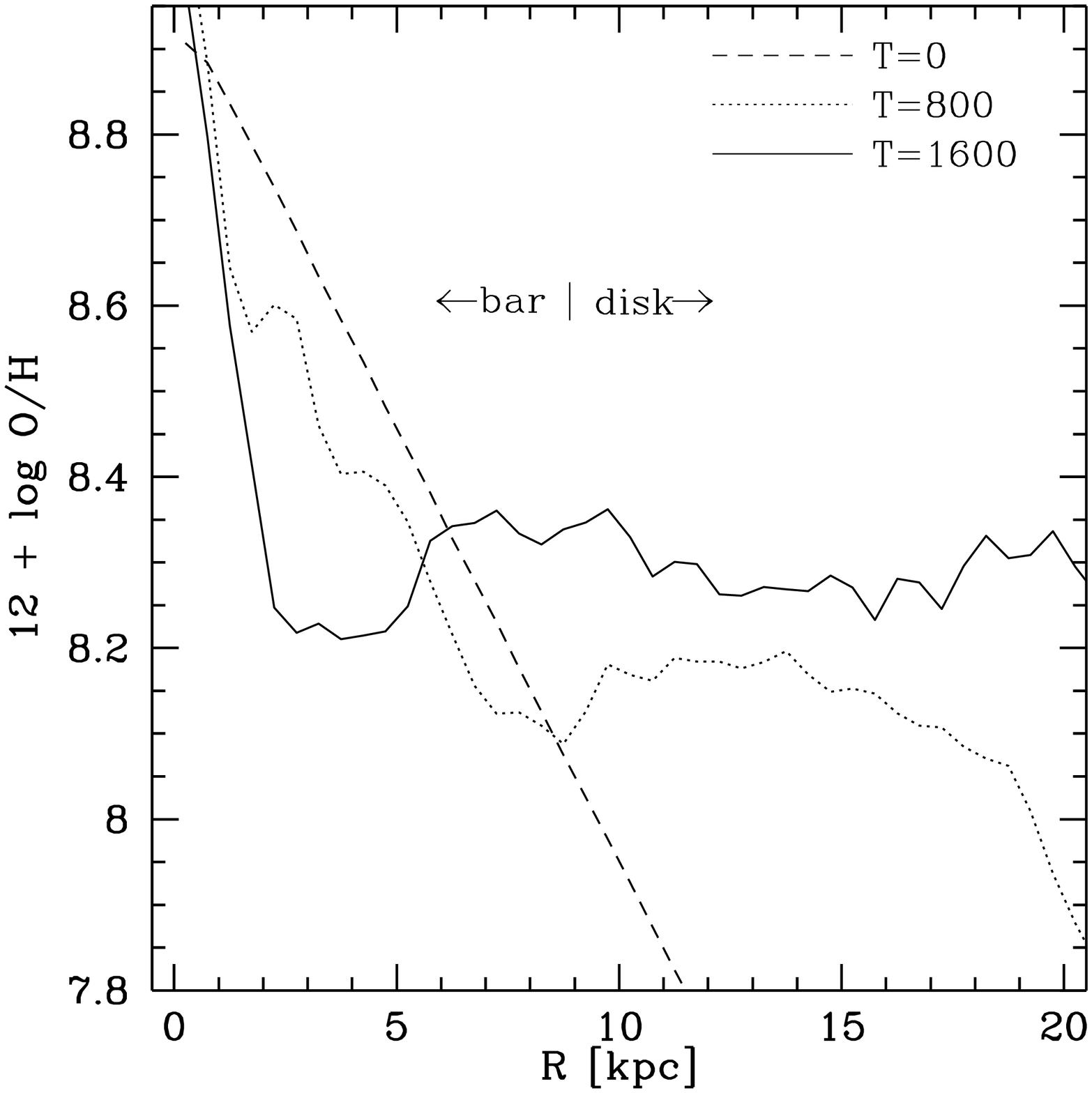}
\caption{{\it Left.} Comparison between stellar and gaseous abundance
profiles in a strongly barred model for the disk region ($T \!=\!
2000$\,Myr). {\it Right.}  Time evolution of the gaseous abundance
gradient. Note the distinct breaks near the end of the bar and in the
disk at $T \!=\! 800$\,Myr.}
\end{figure}

%----------------------------------
\subsection{Bars: Consequences for Gaseous Abundances Profiles}
The previous section clearly tells us that bars should deeply
influence the chemical evolution of galaxies. In particular, one
expects a quick radial homogenization of the cold or warm gas, which
leads to a {\it flattening} of any abundance gradient.  This
intuition is nicely confirmed both by observations (Pagel \& Edmunds
1981; Vila-Costas \& Edmunds 1992; Martin \& Roy 1994; Roy 1996) and
numerical simulations (Friedli et al. 1994; Friedli \& Benz 1995;
Martinet \& Friedli 1997).  The most recent studies also found that
the degree of flattening is directly proportional to the bar strength
(see Martin, this volume).  After 1--2\,Gyr, the gaseous abundance
gradient clearly becomes flatter than the stellar one in the disk
region (Fig.~4, left), and the gas phase is typically over-abundant by
$\sim 0.2-0.4$\,dex with respect to the averaged stellar abundance.

Another recent discovery, mainly due to the boost of measurable
H\,{\sc ii} regions per galaxy, concerns the presence of {\it
breaks} in the radial abundance profiles (Zaritsky et al. 1994; Martin
\& Roy 1995; Roy \& Walsh 1997), i.e. the gradient appears relatively
steep in the bar region, whereas it is very shallow in the disk region
(``steep-shallow break'').  Numerical models indicate that such a
feature is typical in young and strong bars (Martinet \& Friedli
1997). The vigorous star formation along the bar major axis is indeed
able to produce enough new elements to compensate the dilution
resulting from the bar-driven gas inflow.  A clear break occurs near
the CR and lasts for about half a Gyr (Fig.~4, right). At this stage,
at a given radius, spiral arms are also enriched by a few dex with
respect to interarm regions.  These abundance fluctuations are then
progressively erased by azimuthal mixing processes.  It is interesting
to note that the abundance scatter in the disk of NGC~3359 is
approximately 0.4\,dex, i.e. twice the usual observed value. This
galaxy is precisely thought to have a very young bar (Martin \& Roy
1995).

Another break could be observed in the outer part of the disk when the
gradient steepens again (``shallow-steep break'').  There, the long
mixing timescale somewhat delays the complete flattening of the
abundance gradient.  Owing to observational difficulties, this
predicted second break has not yet been detected.

The central starburst generated by the bar formation may also produce
a {\it metal-rich core} (Friedli et al. 1994). The determination of
the very central abundances is very difficult and uncertain.
Circumnuclear ones are however more reliable; they are generally
relatively high, but very similar in barred and unbarred galaxies
(Storchi-Bergmann et al. 1996).  This apparent non-enrichment of the
nucleus of barred galaxies might be an indication of metal outflows in
superwinds (e.g. Heckman 1997). These metals could thus either be
locked in the hot gaseous phase, or have been ejected from the galaxy.
They might also have been swallowed by a central supermassive black
hole, if present.

%----------------------------------------------------------------------
\section{Summary}
In strongly barred galaxies, dynamically-induced abundance profiles
are functions of both time and space, and major characteristics can
schematically be summarized, using the definitions of Eq.~1, as
follows:
\vskip -4truemm
\begin{table}
\begin{center}
\begin{tabular}{rcccl}
  {\it Stars:}
& $|{\rm A}_{R}^{\rm bar}| > |{\rm A}_{R}^{\rm disk}|$, 
	${\rm A}_{R}^{\rm CR} \approx 0$
& ${\rm A}_{\phi}^{\rm gal} \approx 0$
& ${\rm A}_{z}^{\rm box} \approx 0$ 
& {\ } \\
  {\it Gas:} 
& $|{\rm A}_{R}^{\rm bar}| > |{\rm A}_{R}^{\rm disk}|$
& ${\rm A}_{\phi}^{\rm gal} \neq 0$
& {\ }
& {\rm (Young bar)} \\
  {\ }
& ${\rm A}_{R}^{\rm gal} \approx 0$
& ${\rm A}_{\phi}^{\rm gal} \approx 0$
& {\ }
& {\rm (Old bar)} \\
\end{tabular}
\end{center}
\vskip -8truemm
\end{table}

%----------------------------------------------------------------------
\section{Conclusion}
Barred dynamics significantly alters stellar and gaseous abundance
profiles in only a fraction of a Hubble time.  Strong bars have both a
{\it direct} influence via very efficient mixing and transfer of
metals, and an {\it indirect} effect via important element production
through the possible enhancement of star formation.

In order to properly and fully interpret the observed abundance
profiles in galaxies, galactic chemical evolution models should
clearly take into account {\it dynamics}, particularly the triaxial
one, in addition to the classical and basic ingredients like yields,
IMF, and star formation rate.

%----------------------------------------------------------------------
\acknowledgments 
I thank L.~Drissen and J.-R.~Roy for a careful reading of the
manuscript and useful comments.  I am indebted to P.~Martin for having
initiated me into the topic of abundance profiles a few years ago!  I
acknowledge the Swiss National Science Foundation (FNRS) for its
support through an ``Advanced Researcher'' fellowship, and the
Universit\'e Laval for its kind hospitality.

%----------------------------------------------------------------------

%----------------------------------------------------------------------
\end{document}